\documentclass[12pt,preprint]{aastex}

\newcommand{\bdv}[1]{\mbox{\boldmath$#1$}}

\def\au{{\rm au}} 
 
\def\kms{{\rm km}\,{\rm s}^{-1}}
\def\masyr{{\rm mas}\,{\rm yr}^{-1}}
\def\kpc{{\rm kpc}}
\def\mas{{\rm mas}}

\def\muas{\mu{\rm as}}

\def\max{{\rm max}}
\def\min{{\rm min}}
\def\rel{{\rm rel}}

\def\eff{{\rm eff}}

\def\e{{\rm E}}
\def\bpi{{\bdv\pi}}
\def\bmu{{\bdv\mu}}

\def\btheta{{\bdv\theta}}

\def\ch{ }
\begin{document}
\title{Masses for Free-Floating Planets and Dwarf Planets}

\author{\textsc{
Andrew Gould$^{1,2}$, 
Weicheng Zang$^{3,4}$,
Shude Mao$^{3,5}$, and
Subo Dong$^{6}$}}
\affil{$^{1}$Max-Planck-Institute for Astronomy, K\"{o}nigstuhl 17, D-69117 Heidelberg, Germany}

\affil{$^{2}$Department of Astronomy, Ohio State University, 140 W. 18th Ave., Columbus, OH 43210, USA}

\affil{$^{3}$Department of Astronomy, Tsinghua University, Beijing 100084, China}

\affil{$^{4}$Corresonding author}

\affil{$^{5}$National Astronomical Observatories, Chinese Academy of Sciences, Beijing 100101, China}

\affil{$^{6}$Kavli Institute for Astronomy and Astrophysics, Peking University, Yi He Yuan Road 5, Hai Dian District, Beijing 100871, China} 

\begin{abstract}

The mass and distance functions of free-floating planets (FFPs)
would give major insights into the formation and evolution of
planetary systems, including any systematic differences between
those in the disk and bulge.  We show that the only way to measure
the mass and distance of individual FFPs over a broad range of
distances is to observe them simultaneously
from two observatories separated by $D\sim {\cal O}(0.01\,\au)$ (to measure
their microlens parallax $\pi_\e$) and to focus on the finite-source point-lens
(FSPL) events (which yield the Einstein radius $\theta_\e$).
By combining the existing KMTNet 3-telescope observatory with a
0.3m $4\,{\rm deg}^2$ telescope at L2, of order 130 such measurements
could be made over four years, 
down to about $M\sim 6\,M_\oplus$ for bulge FFPs and
$M\sim 0.7\,M_\oplus$ for disk FFPs.  The same experiment would
return masses and distances for many bound planetary systems.
A more ambitious experiment, with two 0.5m satellites (one at L2
and the other nearer Earth) and similar camera layout but in the
infrared, could measure masses and distances of sub-Moon mass objects,
and thereby probe (and distinguish between) genuine sub-Moon FFPs and 
sub-Moon ``dwarf planets'' in exo-Kuiper Belts and exo-Oort Clouds.

\end{abstract}


{\section{{Introduction}
\label{sec:intro}}

The mass and distance distributions of free-floating planets (FFPs)
are crucial diagnostics of planet formation and evolution.  Low 
(``planetary'') mass objects, $M<13\,M_J$,
can in principle either form by gravitational collapse in situ or
be expelled from planetary systems after forming from a protoplanetary
disk.  However, the 12 FFP candidates discovered to date 
(\citealt{mroz17,kb172820}, and references therein) have masses that
are either $M\la 0.2\,M_J$ or $M\la 8\,M_\oplus$, if they reside
in the Galactic bulge or the Galactic disk, respectively.  These
mass ranges are far too small for formation by gravitational collapse,
so they must have formed within protoplanetary disks.

In principle, it is possible that some or all of these FFP candidates
are actually wide-orbit planets\footnote{There are multiple possible
paths to wide-orbit planets, including in situ formation, smooth-pumping
or violent ``relocation'' during or after the planet-formation phase,
or late-time adiabatic orbit expansion due to mass lose.  Hence, if the
FFP candidates prove to be wide-orbit planets, their detailed study
will be an important probe of all these processes.}, 
whose hosts do not leave any
signature on the apparently single-lens/single-source (1L1S) microlensing
light curves from which they are discovered.  This issue will be settled
by late-time adaptive optics (AO) imaging, after the source and the
putative host are sufficiently separated to be resolved.  This will
be possible for all 12 at AO first light on 30m telescopes (roughly 2030), 
and for some a few years earlier \citep{kb172820}.  Until that time,
we will not know that FFPs actually exist.  Nevertheless, \citet{kb172820}
argue that most of these FFP candidates are likely to be true FFPs (rather
than wide-orbit planets), and we will adopt that perspective here.

FFPs that have masses well below those of the typical perturbers
behave as test particles.  Therefore the mass function of FFPs in this
regime should be similar to that of the bodies in the general
region of these perturbers, i.e., at 1--3 times the snow line,
where most gas giants and ice giants reside.  This will already
provide one major diagnostic of conditions in the protoplanetary
and post-protoplanetary disk.  Second, one expects that this
distribution will be strongly cut off as the mass of the FFPs 
approaches that of the perturbers, so that they no longer behave
as test particles.  Hence, this cut-off mass will be another key
diagnostic.  Finally, the FFP frequency and mass function may differ
in different environments, particularly between those in the
bulge and the disk, which would provide insight into the different
planet-formation processes in these two regions.  More generally, there
could be features in the mass and/or distance distribution that we
cannot anticipate today in the absence of data.

Whether the FFP mass and distance distributions are measured
this decade, this century, or this millennium, the method will be
the same: two wide-field telescopes, separated by $D\sim{\cal O}(0.01\,\au)$
will observe at least several square degrees of the Galactic bulge
for an integrated time of at least several years.

The first reason for this is that FFPs in the $M\la 0.2 M_J$ regime
can only be studied by gravitational microlensing.  They are unbound,
and so they cannot be detected via their gravitational effect on
any other object, nor by their blocking light from any other object.
Some FFPs may emit thermal radiation due to heat trapped from formation
or violent encounters.  However, the only ``guaranteed'' source of
thermal emission (which is what is required for a survey based on
homogeneous detections) is radioactive decays.  For Earth, with
its ``typical'' age of 4.5 Gyr,  this amounts to $2\times 10^{20}$
ergs/s, or $5\times 10^{-14}\,L_\odot$, which (for a black body)
would be emitted at
$T\sim 29\,$K, with a {\ch bolometric magnitude $M_{\rm bol} = 37.9$,
if Earth were ``free''.  This would correspond to $m_{\rm bol} = 52.5$}
for an Earth-like FFP in the Galactic bulge.
Therefore, FFPs are effectively ``dark''.  Hence, their only detectable
effect is that they focus light from more distant stars.  Indeed,
a dozen FFP candidates have been detected via this route.

Second, once detected, the only way to determine the mass of a dark, isolated 
object is to measure both its angular Einstein radius $\theta_\e$
and its microlens parallax $\pi_\e$,
\begin{equation}
\pi_\e\equiv {\pi_\rel\over\theta_\e};
\qquad
\theta_\e\equiv \sqrt{\kappa M\,\pi_\rel};
\qquad
\kappa\equiv {4G\over c^2\,\au}\simeq 8.14\,{\mas\over M_\odot}.
\label{eqn:pie_thetae}
\end{equation}
Here, $\pi_\rel\equiv \au(D_L^{-1}-D_S^{-1})$ 
is the lens-source relative parallax, which for
bulge lenses is of order\footnote{For the typical case that
$(D_S+D_L)/2= 8\,\kpc$ (the approximate distance to the bulge) and
$(D_S-D_L)= 1\,\kpc$ (the approximate thickness of the bulge).}
 $\pi_\rel \sim 16\,\muas$.

There are only two ways to measure $\pi_\e$: simultaneous photometry
from two observatories during the event \citep{refsdal66}, or, 
photometry from a single accelerated platform during the event
\citep{gould92}.  The shortest FFP events will have their timescale
set by the source crossing time $t_* = \theta_*/\mu_\rel$
(which is of order an hour for main-sequence sources), rather than
their Einstein crossing time $t_\e = \theta_\e/\mu_\rel$. 
Here $\theta_*$ is the angular size of the source and $\mu_\rel$
is the lens-source relative proper motion.  Hence, to measure $\pi_\e$
using the second method, the orbital period of the accelerated platform
should be of order an hour, which is well matched to low-Earth orbit
\citep{honma99}.
However, for bulge lenses, the projected size of the source is \citep{gould13},
\begin{equation}
\tilde R_* = \rho \tilde r_\e = 
\rho{\au\over \pi_\rel/\theta_\e}=
\au{\theta_*\over \pi_\rel} = 
880\,R_\oplus\biggl({\theta_*\over 0.6\,\muas}\biggr)
\biggl({\pi_\rel\over 16\,\muas}\biggr)^{-1},
\label{eqn:rs}
\end{equation}
where $\tilde r_\e\equiv \au/\pi_\e$ is the Einstein radius projected
on the observer plane and $\rho\equiv \theta_*/\theta_\e$ is the
angular source radius scaled to the angular Einstein radius.
Hence, for small bulge FFPs, there would be essentially no parallax
signal as the observatory orbited Earth.  Thus, the only method of
measuring $\pi_\e$ (and so masses) for a broad range of FFPs, in both the disk
and the bulge, requires two well-separated observatories.

In principle, there are several methods of measuring $\theta_\e$ for
dark objects.  For example, in astrometric microlensing, the centroid
of microlensed light deviates from that of the source by
$\Delta\btheta = \delta\btheta/[(\delta\theta/\theta_\e)^2 + 2]$,
where $\delta\btheta$ is the lens-source separation vector
\citep{my95,hnp95,walker95}.  However, first, this requires measuring
astrometric deviations $\theta_\e/\sqrt{8}\rightarrow 0.35\,\muas$ for
the smallest $\theta_\e\sim 1\,\muas$ under consideration, which
is set by the smallest accessible sources (corresponding to
$M=2.6\,M_\oplus$ for $\pi_\rel=16\,\muas$ bulge lenses and
$M=0.33\,M_\oplus$ for $\pi_\rel=125\,\muas$ disk lenses\footnote{Represented
by $D_L = 4\,\kpc$ and $D_S = 8\,\kpc$.}).
Second,
it requires an alert to the astrometric telescope on timescales
$<t_\e\sim 1\,$hr.  For a relatively precise measurement, a dozen
100 nas (i.e., $3\,\sigma$)
measurements should be acquired within a few hours on an $I\sim 20$
target.  A second method would be to resolve the two images using
interferometry \citep{delplancke01,kojima1b}.  
However, the $2\,\muas$ resolution
required is about 1000 times better than current interferometers,
which only work on targets that are about 1000 times brighter.  In addition,
this would require alerting these massive instruments on timescales
$< t_\e = 1\,$hr.  Thus, the only practical method is to observe events
for which the lens passes directly over the face of the source, leading
to a light curve that is described by four parameters $(t_0,u_0,t_\e,\rho)$,
where $t_0$ is the time peak and $u_0$ is the impact parameter
(normalized to $\theta_\e$) 
\citep{gould94a,witt94,nemiroff94}.  Then $\theta_\e = \theta_*/\rho$,
where $\theta_*$ can be determined using standard techniques\footnote{
In brief, the intrinsic source color and magnitude 
[e.g., $[(V-I),I]_{0,s}$]
are determined from the observed offset 
$\Delta[(V-I),I] = [(V-I),I]_s -[(V-I),I]_{\rm cl}$ on a color-magnitude
diagram of fields stars, together with the known intrinsic position
of the red clump $[(V-I),I]_{0,\rm cl}$ \citep{bensby13,nataf13}.  Using
an empirical color/surface-brightness relation (e.g., \citealt{kervella04}),
often after transforming to $(V,K)$ bands using color-color relations
(e.g., \citealt{bb88}), one then derives the surface brightness
and so solves for $\theta_*$ using the physical relation $F = \pi S\theta_*^2$,
where the source flux $F$ and the surface brightness $S$ are on the
same system.
} \citep{ob03262}.
Such transits occur with probability
$\rho$, which is of order $10^{-2}$ -- $10^{-3}$ for typical microlensing
events.  However, because $\theta_\e$ is small for FFP candidates, $\rho$
is much larger.  Indeed, half of FFP candidates found to date have
such finite-source point-lens (FSPL) light curves, and hence
$\rho$ measurements \citep{kb172820}.

In brief, the only conceivable route to measuring the mass and distance
distributions of FFP candidates over a broad range of distances
is by synoptic observations from two
observatories that are separated by many Earth radii.  

Here, we map the path toward making these measurements.  We begin
by further quantifying the two requirements described above, i.e.,
to measure $\pi_\e$ from a pair of observatories and to measure $\theta_\e$
from FSPL events.  Next, we discuss specific possible implementations,
beginning with those that can take advantage of existing resources
and moving toward more complex and difficult experiments.  We show
that the mass function of the ``known class'' of FFPs
\citep{mroz17,kb172820} can be measured in the ``near'' (5--10 year)
future.  A more ambitious, but already feasible, experiment could
study sub-Moon ``dwarf planet'' FFPs, as well as similar objects
that remain bound in exo-Kuiper Belts (exo-KBOs) and exo-Oort Clouds
(exo-OCOs).  We also comment on the
additional microlensing science that would be returned by these
efforts.

{\section{{Microlens Parallax Requirements}
\label{sec:parallax}}

We begin by analyzing the requirements for making the measurements
in a very general way before considering specific implementations.

The first general requirement is that the lens and source be sufficiently
separated in the Einstein ring that the light curves differ enough
to allow a parallax measurement.  This places a lower limit on the projected
separation of the two observatories $D_\perp$.
We designate the vector separation of the two observatories as ${\bf D}$,
which at any given time yields a projected separation on the sky
${\bf D}_\perp$.  In fact, we will mostly be concerned with the
magnitude of this 2-D vector, i.e., $D_\perp$.  The ratio of this separation
to the projected radius of the source (similar to Equation~(\ref{eqn:rs}))
is
\begin{equation}
{D_\perp\over \tilde R_*} = {D_\perp\over \au}\,{\pi_\rel\over\theta_*}
=0.27\,\biggl({D_\perp\over 0.01\,\au}\biggr)
\biggl({\pi_\rel\over 16\muas}\biggr)
\biggl({\theta_*\over 0.6\muas}\biggr)^{-1}.
\label{eqn:rs2}
\end{equation}

We have normalized Equation~(\ref{eqn:rs2}) to $\theta_* = 0.6\,\muas$,
which is the source radius of the most common type of ``reasonably bright''
FSPL FFP event (as we will discuss in more detail in Section~\ref{sec:fspl}).
And we have also normalized $\pi_\rel$ to that of a typical bulge lens,
which are the most challenging FFPs.  With the fiducial parameters of
Equation~(\ref{eqn:rs2}), the peak times $t_0$ would differ by $0.27\,t_*$
assuming that the lens-source motion $\bmu_\rel$ were aligned with
${\bf D}_\perp$.  And the two trajectories would be displaced by
$0.27\,\theta_*$ if $\bmu_\rel$ and ${\bf D}_\perp$ were orthogonal.
Because these quantities can easily be measured to 1/10 of these values
with reasonable data, this separation is quite adequate.  

The second requirement is that the source trajectories as seen by each
observatory should come within the Einstein radius of the lens.
Otherwise one will obtain only a lower limit on this separation, and hence
only a lower limit on $\pi_\e$.  Of course, {\ch for events with $\rho>1$
one could measure this offset up to separations 
$\sim \theta_* = \rho\theta_\e$, and,
with sufficiently good data,
one could measure it up to $\sim 2\theta_\e$ (or more) even for events 
with $\rho\la 1$.}.  
However, in the limiting cases that define this criterion,
measurement at one Einstein radius will be challenging.  And we also
note that if $\bmu_\rel$ were perfectly parallel to 
${\bf D}_\perp$, then both trajectories would have the same impact
parameter, regardless of the magnitude of $D_\perp$.  However, the criterion
should be set by the general problem of detectability, not special cases.
That is, the magnitude of the normalized separation, 
\begin{equation}
\Delta u = |\Delta {\bf u}| = 
\bigg|{{\bf D}_\perp\over \tilde r_\e} \bigg|
= {D_\perp\over \au}\,\sqrt{\pi_\rel\over \kappa M},
\label{eqn:deltau}
\end{equation}
should be $\Delta u < 1$.

An important aspect of the experiment is that it should be sensitive
to lenses of the same mass in both the disk and the bulge.  
Equation~(\ref{eqn:deltau}) shows that at fixed lens mass,
$\Delta u\propto \sqrt{\pi_\rel}$.  Hence, for sufficiently large $\pi_\rel$
the source as seen from the second observatory will be ``driven out'' of
the Einstein ring.  However, if we consider the smallest bulge lenses
from the example above,
with $\theta_\e=\theta_* = 0.6\,\muas$ (and therefore
$M=\theta_\e^2/\kappa\pi_\rel= 0.9 M_\oplus$), then this condition will
be met provided that $\pi_\rel < (16\,\muas)/0.27^2 = 220\,\muas$,
corresponding to $D_L>2.9\,\kpc$.  After taking account of the fact
that at somewhat larger $\pi_\rel$ there will still be many measurements
due to non-orthogonal trajectories,  a very broad range of distances
will be included even for the case of the most difficult 
mass for bulge detections.

To review, because it is possible to make a parallax measurement when
the offset in Einstein ring $\Delta u$ is much less than the normalized
source size 
$\Delta u/\rho=D_\perp/\tilde R_*\ll 1$, 
it is also possible to keep the lens-source
separation inside the Einstein ring for a broad range of {\ch distances:
$\pi_{\rel,\rm bulge} < \pi_{\rel} < (\tilde R_*/D_\perp)^2\pi_{\rel,\rm bulge}$.}

{\section{{FSPL Requirements}
\label{sec:fspl}}

In one sense, the FSPL requirement is exquisitely simple: the lens must
transit the source, i.e., come within $\theta_*$ of its center.  However,
the range of properties of potential sources is enormous, and any 
concrete experimental FSPL-survey design must focus on some subset or subsets.
For example, \citet{kb192073} focused on giants.  Moreover, the FSPL
component of a survey that incorporates parallax, must take account
of the constraints arising from the parallax measurement 
(see Section~\ref{sec:parallax}).

Before reviewing the characteristics of the source population, we note
that the event rate (as a function of lens mass) for FSPL events
is very different from the microlensing event rate.  For an individual source,
with radius $\theta_*$, these are,
\begin{equation}
\Gamma_{\rm FSPL}(M) = 2\theta_*\int_0^{D_s} d D\, D^2 n(M,D)\langle\mu(D)\rangle,
\label{eqn:fsplrate}
\end{equation}
and
\begin{equation}
\Gamma_{\rm event}(M) = 2\int_0^{D_s} d D\, D^2 n(M,D)\langle\mu(D)\rangle
\theta_\e(M,D) = 
2\int_0^{D_s} d D\, D^2 n(M,D)\langle\mu(D)\rangle\sqrt{\kappa M\pi_\rel},
\label{eqn:eventrate}
\end{equation}
where $n(M,D)$ is the number density of lenses with mass $M$ and
distance $D$, and where $\langle\mu(D)\rangle$ is the mean 
lens-source relative proper motion of these lenses.  Due to the last
factor in Equation~(\ref{eqn:eventrate}), more massive lenses and more
nearby lenses are heavily favored relative to their number density in
the overall event rate, which is not true of the FSPL rate.  From the
standpoint of studying FFPs, this FSPL bias toward low mass objects is
obviously good: if there are really 5--10 times more super-Earth FFPs
than stars \citep{mroz17,kb172820} then there will be 5--10 times more
FSPL FFP events than FSPL stellar events.  However, from the
standpoint of probing a broad range of distances (and so a broad range
of environments), this bias is somewhat troubling.  Due to the low
surface density of disk stars, they contribute a minority of all
events, even with their $\sqrt{\pi_\rel}$ advantage shown in
Equation~(\ref{eqn:eventrate}).  This shortfall will now be multiplied
by a factor $[(125\,\muas)/(16\,\muas)]^{1/2}=2.8$, which will be an
important consideration further below.  Finally, we recall from
Section~\ref{sec:parallax} that at fixed mass, low $\pi_\rel$ (e.g.,
bulge) lenses drop out of the sample due to the difficulty of
measuring their microlens parallax.  Thus, any survey design strategy
must take account of both the intrinsically low rate of disk FSPL events
and the suppression of low-mass bulge events in the process of their
 parallax measurement.

To frame the issues of survey design, we first make a rough estimate
of the event rate from G dwarf sources using the \citet{holtzman98} luminosity
function, which we first multiply by a factor two because the density
of sources and lenses is about two (or more) times higher in the
best microlensing fields 
(\citealt{nataf13}; D.\ Nataf 2019, private communication).
We ignore disk lenses because, as just discussed, they are a 
numerically minor (though scientifically
very important) addition to the overall rate.  The rate per unit area is
\begin{equation}
{d\Gamma_{\rm FSPL/G}\over d\Omega}=2\, \langle \theta_*\rangle N_{\rm FFP}N_{\rm G}
\langle \mu\rangle \rightarrow 
2\times 0.5\,\muas\,
{5\times 10^5\over {\rm arcmin^2}}{3\times 10^3\over {\rm arcmin^2}}\,
6.5\,\masyr
= {9.8\,{\rm yr}^{-1}\over {\rm deg^2}},
\label{eqn:gdwarf}
\end{equation}
where we have adopted $\langle \theta_*\rangle=0.5\,\muas$ as the mean
radius of G dwarfs, defined as $3.5<M_I<5$.
We estimate $N_{\rm G}$, the surface density of G dwarfs, by doubling
the number within $3.5<M_I<5$ in Figure~5 of \citet{holtzman98}.
We extrapolate this diagram to estimate the surface density of
stars as $5\times 10^4/{\rm arcmin^2}$, then double this number
to consider a better microlensing field, and then multiply by five
based on the 5:1  FFP/star ratio estimated by \citet{mroz17}.
We approximate the bulge proper-motion distribution as an isotropic
Gaussian with dispersion $\sigma = 2.9\,\masyr$ based on experience
with {\it Gaia} proper-motion data in many high event-rate fields.
This functional form implies $\langle \mu\rangle = (4/\sqrt{\pi})\sigma$.
See Appendix.

Next we repeat this calculation for three other brighter classes
of stars, 
turnoff/subgiants ($2< M_I< 3.5$),
lower-giant-branch ($0.5< M_I< 2$), and
upper-giant-branch+red-clump ($M_I< 0.5$).  For the four classes,
we adopt surface-density ratios
(1.000, 0.267, 0.027, 0.025) and cross {\ch sections
$2\langle \theta_*\rangle=(1.0,2.4,9.0,14.0)\muas$.
The product of these factors is $(1.00,0.64,0.24,0.35)$.  Hence,}
scaling to Equation~(\ref{eqn:gdwarf}), these yield respective rates
\begin{equation}
{d\Gamma\over d\Omega}\Biggl(\matrix{\rm G\ dwarfs \cr 
\rm Turnoff/Subgiants \cr \rm Lower\ Giants \cr \rm Upper\ Giants
}\Biggr)=
\Biggl(\matrix{9.8\cr 6.2\cr 2.3\cr 3.4}\Biggr)
{{\rm yr}^{-1}\over {\rm deg^2}}.
\label{eqn:alltype}
\end{equation}

The first point to note regarding Equation~(\ref{eqn:alltype}) is that
there can be a large number of potential FFP mass measurements,
provided that some or all of these regimes can actually be probed.  
There are about 
$10\,{\rm deg}^2$
of high event-rate fields in the southern bulge that have modest
extinction, $A_I\la 2$, for which the G-dwarf limit 
would require $I_{\rm lim}\sim 21.5$.
Now, it is certainly not possible to properly characterize magnification
$A\sim 2$ events on $I=21.5$ sources from any current ground-based
surveys, so the simplest implementation of this approach (coupling
a new observatory orbiting at L2 with existing ground-based surveys)
could not reach this limit.  

However, the defining target of the first survey would be the
bulge analogs of the $\theta_\e\sim 6\,\muas$ disk FFP population that 
has already been detected, i.e., {\ch with 
$\theta_\e=(16/125)^{1/2}6\,\muas=2\,\muas$.  To be sensitive to
a broad range of bulge $\pi_\rel$, we adopt a more conservative
fiducial value of $\theta_\e=1.5\,\muas$.  For these,}
$\rho = \theta_*/\theta_\e= 0.33$ and so the peak magnification is
$A_\max = \sqrt{1 + 4/\rho^2}\rightarrow 6$, implying a difference-star
magnitude of $I=19.8$.  Such an event probably could not be reliably
recognized in ground data alone.  
However, if the L2 telescope had substantially
better data, and in particular could determine $t_\e$ and $\rho$, then
the fit to the ground-based light curve would be highly constrained
(the reverse of the situation considered by 
\citealt{gould95,boutreux96,gaudi97}).
The situation would be substantially
better for G dwarfs in the middle of the distribution, i.e., a factor
2 brighter.

We now turn to the opposite extreme: giant sources.
The same bulge super-Earth discussed above
would magnify only a small part of a clump giant's $(\theta_*=6\,\muas)$
surface, implying 
$A_\max=1.12$ and so $I_{\rm diff} = 18.8$, i.e., a magnitude brighter
than the G-dwarf case.  The background (due to the giant itself) is
higher, but this is overall a secondary effect.

The lower-giant branch stars have similar color, so similar surface
brightness.  Because only a portion of their surface would be magnified
by a $\theta_*=1.5\,\muas$ lens, the difference star would have similar
brightness $I_{\rm diff} = 18.8$.  Moreover, the source itself would
generate less background noise.

The best case would be the turnoff/subgiants because they have higher
surface brightness.  For example, for $M_I=3$ and $\theta_*=1.2\muas$,
$I_{\rm diff} = 18.6$. That is, all four classes in Equation~(\ref{eqn:alltype})
could potentially contribute to FFP detections, although it will still
be necessary to examine the integrated measurement process as a whole. 

In brief, there is a known population of 12 FFPs, of which 11 are
likely due to super-Earths, mostly in the disk \citep{kb172820}, with
five of these 11 having measured $\theta_\e\sim 6\,\muas$.  If there
are analogs of these objects in the bulge (with $\theta_\e\sim 1.5\,\muas$),
then none have been detected in current surveys, and the sensitivity of these
surveys to such objects is limited\footnote{OGLE-2016-BLG-1928
has $\theta_\e\sim 1\,\muas$, but it is almost certainly a much
lower mass object that lies in the disk \citep{ob161928}.  The fact that
it was detected shows that current surveys have some sensitivity
to bulge analogs of the detected FSPL events, although it is weak.}.
However, even at the adopted $M_I\sim 5$ threshold, ground
surveys could marginally characterize the light curve generated by
such putative bulge super-Earths, provided that $(\rho,t_\e)$ were
well determined from space.  This would permit a marginal mass measurement
at this threshold.  Mid-G dwarf and brighter sources
would yield substantially better results.

{\section{{KMTNet + L2 Satellite (KMT+L2)}
\label{sec:kmt+l2}}

In this and the next section, we will consider two of the
many possible two-observatory scenarios that could probe the FFP
mass and distance functions.  We begin this section by motivating
why combining the KMTNet survey \citep{kmtnet}
with an L2 satellite (KMT+L2) should
be one of those subjected to review\footnote{KMTNet 
combines three telescopes, located in Australia (KMTA), 
Chile (KMTC) and South Africa (KMTS), each with a 1.6m telescope, equipped
with an 18k$\times$18k camera spanning a $4\,{\rm deg}^2$ field.
Of some practical import in the present context, the telescopes are on
equatorial mounts, and the field is oriented on equatorial coordinates.
See, e.g., Figure~12 of \citep{eventfinder}.}.

First, KMT+L2 is an intrinsically cheap option.  The satellite requirements
are limited by the fact that whatever FSPL events that it might detect that
are ``inaccessible'' to KMTNet (in the sense that they cannot be 
characterized with ground-based data even if $t_\e$ and $\rho$ are 
known infinitely well from space) are useless.

Second, such a low-requirement satellite could be built very
quickly, while KMTNet is still in operation (or could be persuaded
to remain in operation).  Thus, it could return results on FFPs
before it is absolutely confirmed that the bulk of the FFP candidates
that have been reported to date are FFPs (rather than wide-orbit planets).

Note that while wide-orbit planets, if they exist, would be just
as interesting and important as FFPs, they do not require such
specialized equipment to measure their mass and distance functions.
The very same 30m AO followup that would prove that the FFP candidates 
have hosts, would also measure the mass and distance of these hosts, while 
second AO epochs would yield the
host-planet separations \citep{gould16,kb172820}.

Therefore, the low cost of KMT+L2 is well matched to the higher
risk that the target population may be non-existent.

Third, by obtaining early results, KMT+L2 could influence the
design of more advanced experiments that would be motivated
by AO confirmation of earlier FFP candidates.  For example, \citet{kb172820}
show that FFP candidates
(OGLE-2016-BLG-1540, 
OGLE-2016-BLG-1928,
OGLE-2012-BLG-1323,
KMT-2017-BLG-2820)
\citep{ob161540,ob161928,ob121323,kb172820}
can be confirmed (or ruled out) as FFPs by 
(2024, 2024, 2027, 2028), respectively.

Fourth, as we will show, any experiment designed to measure masses
and distances of FFPs will automatically return these same measurements
for a large fraction of bound-planet lenses in its field of view.
Such measurements will remain of exceptional interest only until the
advent of 30m AO, at which point such mass and distance measurements
will generally be possible after wait times of 3--5 years.  The exception
is that two-observatory experiments will also yield masses and
distances for dark (e.g., brown dwarf, white dwarf) hosts, whereas
AO followup will not.

As a specific example, we will consider a 0.3m optical telescope in L2,
equipped by a 18k$\times$18k camera.  This choice is partly motivated
by the actual design of a planned multi-telescope satellite 
(Earth 2.0 Transit Survey Mission), which
will mainly be utilized for transits, but which could include a microlensing
telescope, and partly because a 0.3m telescope
would yield photometry that is significantly (but not dramatically)
better than KMTNet at the faint end, 
in accordance with the first motivating point
given above.

We will assume a throughput similar to KMTNet and a filter
similar to $I$-band as well.  These imply a FWHM = $0.67^{\prime\prime}$
and photometric zero point of $I_{\rm zero}=26.75$ for a nine-minute
exposure, i.e., 200 photons from an $I=21$ difference star.

We now consider a specific implementation with a
$2^\circ\times 2^\circ$ field of view and $0.40^{\prime\prime}$ pixels,
i.e., identical to KMTNet.
The camera would be centered and oriented to exactly match
the KMT observations.  The center would be at about 
$(l,b)=(+1.0,-1.8)$ if the KMTNet cameras can be rotated to Galactic
coordinates and about $(l,b)=(+1.0,-2.1)$ if they cannot.
 The blue dots in Figure~8 of \citet{kb181292}
show published planetary microlensing events from 2003--2010, a period
when microlensing survey cadences were adequate to find events over a
broad area, but mostly could not characterize planetary perturbations
by themselves.  Rather, planets were mostly found by targeted follow-up
observations of these events \citep{gouldloeb}.  
The distribution is quite broad over the
southern bulge, although it does favor regions that are closer to the
plane.  Hence, the huge concentration of planet discoveries centered
on $(l,b)=(+1.0,-1.8)$ from subsequent years is mainly a product of the
fact that much higher-cadence surveys concentrated on these areas.
Nevertheless, KMTNet's choice to concentrate on this area with its
highest cadence fields (red field numbers) does reflect the highest
intrinsic event rate as determined using the method of \citet{poleski16}.

At $(0.67/0.40)=1.7\,$pixels/FWHM, the images would be slightly subsampled,
but still much better sampled than for $3.6\mu$m observations on {\it Spitzer}:
0.9 pixels/FWHM.  And the photometry would benefit from the much more
uniform pixel response function characteristic of optical CCDs.  Hence,
it is plausible that photon-limited photometry could be achieved.  
Considering the point spread function (PSF) of 
$\pi\,{\rm FWHM}^2 = 1.4\,{\rm arcsec}^2$,
there would be about 1.5 G dwarfs (similar to the target source population)
per PSF.  Hence, an $A=2$ magnification event (of an $M_I=4.5$, $A_I=2$
star) would have a 200 photon
signal and $\sqrt{200\times 2.5}\sim 25$ photon noise (assuming good
read noise, etc), i.e., a signal-to-noise ratio (SNR) of 8.  With 10
such exposures over a typical $2\,t_*=1.6\,$hr event, there would be a
clear detection and reasonable characterization of extreme 
$\theta_\e= \theta_*\sim 0.5\,\muas$ FFPs.  However, because this
same event could barely be ``detected'' from the ground\footnote{
See KMT error bars near $I\sim 20.2$ in Figure~\ref{fig:lc0}.} (even if one
knew from the space observations where to look), there would be no
parallax measurement.  Still, the ``routine'' detection of such
small-$\theta_\e$ FSPL events would be of considerable interest.
By contrast, only one such $\theta_\e< 1\,\muas$ event has been
detected in 10 years of OGLE-IV data \citep{ob161928}.

Because $\theta_*=0.5\,\muas$ is already in the $\rho\ga 1$ limit for
which the excess flux is basically just twice the area of the Einstein ring
times the surface brightness, the brighter three classes have
similar excess fluxes.

On the other hand, as discussed in Section~\ref{sec:fspl},
for putative bulge super-Earth FFPs of 
$\theta_\e= 3\,\theta_*\sim 1.5\,\muas$,
the satellite would yield excellent characterization and, based on the
resulting $(\rho,t_\e)$ measurements, the ground light curve could
be well characterized.  See {\ch Figures~\ref{fig:lc0} and \ref{fig:lcerr}.}

However, the true rate of measurements would be well below that implied
by Equation~(\ref{eqn:alltype}) simply because these require simultaneous
observations.  While the L2 observations could be carried out
continuously (apart from a short window when the Sun passes through
the bulge),  the combined three KMTNet telescopes can observe a given bulge
field 49\% (after taking account of a 3\% overlap between KMTS and KMTC)
of the year due
to the annual and diurnal cycles.  See Figure~\ref{fig:kmt}.
For about 1/4 of this 49\%
(averaged over the 3 telescopes), bad weather or high background would
prevent useful observations, leaving about 37\%.  For about 30\% of this
remaining time, the projected separation would be $D_\perp < 0.5 D$ due to 
the alignment of Earth, L2, and the bulge near 
(June 20)$\pm$(1 month).  This would
degrade parallax measurements for small $\pi_\rel$, in some cases critically.
We estimate that KMTNet and the L2 satellite would be able to work together
about 28\% of the year.  Equation~(\ref{eqn:alltype}) then implies that
a 4 year mission would make mass/distance measurements for about 100
bulge super-Earth FFPs.  There are, intrinsically, about 5 times fewer 
corresponding disk FFPs.  However, due to their roughly 8 times larger $\pi_\rel$
(and corresponding $\sim 2.8$ times larger $\theta_\e$), they
are hardly affected by the contraction of $D_\perp$ near opposition.
Moreover, the peak magnification for FSPL events 
is more than 1 magnitude greater, meaning
that 1 additional magnitude from the Holtzman luminosity function is
accessible.  Hence, we estimate roughly 30 measurements of disk FFPs
from the same population.  In addition, it is plausible that the
FFP mass function rises toward lower (e.g., Earth and Mars) masses,
in which case the experiment would be sensitive to those as well, but only
in the disk.

Table~\ref{tab:telescopes} summarizes the adopted properties of the
KMT+L2 system, and Table~\ref{tab:sources} summarizes the prospective
FFP detections as a function of source type and lens population.

{\subsection{{Source Color Measurements}
\label{sec:color}}

The determination of $\theta_\e$ requires that the source color
be measured, or at least accurately estimated.  The best way to do
this is to take data in two bands during the event, and either
fit them both to a common model, or just perform linear regression
on the fluxes.  In the present case, the only source of data in a second
band will be KMT $V$-band data.  For the marginal events just described,
with difference magnitudes $I_{\rm diff}=21$, the $V$-band difference magnitudes
will be $V_{\rm diff}=I_{\rm diff}+ (V-I)_0 + E(V-I)\rightarrow 
I_{\rm diff}+ 2.2$ for $(V-I)_0=0.7$ and  $E(V-I)=1.5$.  This will not
be measurable in the most extreme cases $I_{\rm diff}=21$, for which
it will be necessary to estimate the color, either from that of the
baseline object (see Section~\ref{sec:csst})
or from the fitted source flux (or even baseline flux)
together with a color-magnitude diagram.  The former method can
work quite well provided that the baseline images are resolved to the
depth of the source flux.  The latter can lead to
errors in $\theta_\e$ (and so $M$)
of $\sim 15\%$ for main-sequence stars and turnoff stars.

However, for the defining targets, bulge super-Earths with 
$\theta_\e\sim 1.5\,\muas$, a turnoff source, 
$M_V=3.7$, $\theta_*=1.2\,\muas$ and $A_V=3.5$, together imply
$V_{\rm diff} = 21.2$.  This corresponds to about 1000 difference
photons in a 75 second exposure, of which there would be a dozen over peak.
Hence, there would be many robust color measurements as well as some
estimated colors with larger error bars.

{\subsection{{{\it CSST} Imaging for Baseline-Object Color and Blending}
\label{sec:csst}}

The {\it Chinese Space Station Telescope (CSST)} is a 2m wide-field 
($1.2\,{\rm deg}^2$) imager with 75 mas pixels,
scheduled for launch in 2024.  There is no filter wheel, but sections
of the focal plane are allocated to various pass bands, including 
SDSS $(g,r,i,z,y)$\footnote{
Other sections are allocated to $u$ and NUV filters, as well as 
to various grisms.}, with FWHM=(60,82,98,123,136) mas.
Hence, the entire KMT+L2 $4\,{\rm deg}^2$ field could
be covered in $griy$ in 200 overlapping pointings\footnote{Given the
specific layout of the detector, complete coverage in $z$ would
require an additional 200 pointings.}, with about
90\% of the area imaged twice in each band.

At this resolution, and at the depth relevant to the experiment,
the field is essentially ``empty'', i.e., just 
$6\,\times 10^{-3}$ G dwarfs per pixel.  In most cases, the event
could be localized to 0.1 KMT/L2 pixels (40 mas) from difference
imaging.  Hence, very few ambient stars would be mistaken for and/or
blended with the source.  That is, the potential blends would be
essentially restricted to companions of the source or lens.  For
cases that the source flux derived from the fit was in tight agreement
with that of the corresponding baseline object, the baseline-object
color would be an excellent proxy for the source color.  In other
cases, one could adopt the color and magnitude of the baseline object,
together with a suitable statistical distribution based on
properties of potential lens and source companions.  This entire procedure could
be rigorously tested on hundreds of high-magnification microlensing
events, for which the $(V-I)$ color will be precisely measured
by KMT.

{\subsection{{Discrete and Continuous Parallax Degeneracies}
\label{sec:discrete}}

\citet{refsdal66} already pointed out that satellite parallax determinations
are subject to a four-fold degeneracy because we infer
$\bpi_\e = \Delta {\bf u}(\au/D_\perp)$ from a measurement of the
offset in the Einstein ring, $\Delta {\bf u}=(\Delta t_0/t_\e,\Delta u_0)$,
from the fit parameters $(t_0,u_0)$ to the ground and satellite light curves.
However, while $u_0$ is a signed quantity, only its modulus can generally
be determined from the light curve of short events.  Thus, there are two
solutions with the source passing on the same side of the lens as seen
by both observatories, $(+,+)$ and $(-,-)$, and two with the source passing
on opposite sides, $(+,-)$ and $(-,+)$.  The two members of each
pair have the same $\pi_\e$ but different directions.  However, the first pair
has smaller $\pi_\e$ than the second: $\pi_{\e,\pm,\pm} < \pi_{\e,\pm,\mp}$.
See Figure~1 from \citet{gould94b}.  For the very short events under 
consideration here, the only way to rigorously
break this degeneracy is to observe the event 
from a third observatory \citep{refsdal66} as was done in the
cases of (OGLE-2007-BLG-224, OGLE-2008-BLG-279, MOA-2016-BLG-290)\footnote{
The first two were terrestrial-parallax measurements from
(Chile, South Africa, Canaries) and (Tasmania, South Africa Israel),
respectively.  The third combined (Earth,{\it Spitzer},{\it Kepler}).
Neither \citet{ob07224} nor \citet{ob08279} explicitly recognized
that the four-fold degeneracy was broken by three observatories,
although \citet{ob07224} did note the consistency of two time delays,
which is the same issue.  Hence, \citet{mb16290} were the first
to explicitly break this degeneracy.
}
\citep{ob07224,ob08279,mb16290}.

The only other path to distinguishing between $\pi_{\e,\pm,\pm}$ and
$\pi_{\e,\pm,\mp}$ is statistical: if $\pi_{\e,\pm,\pm} \ll \pi_{\e,\pm,\mp}$
(i.e., $\Delta u_0\ll u_0$) then the latter solution requires
fine tuning (J.\ Rich, circa 1997, private communication;
\citealt{21event}).  In fact, the ``Rich Argument'' factor appears
naturally as a Jacobian within a standard Bayesian analysis
\citep{gould20}.

For KMT+L2 FFP parallax measurements, the ``Rich argument'' 
will often prove applicable to giant-source events.
For example, according to Equation~(\ref{eqn:rs2}), for a 
lower-giant-branch source ($\theta_*=3\,\muas$) and a bulge lens
$\pi_{\rm rel}=16\,\muas$, the offset between the two observatories
will be 0.054 source radii.  Hence, for $u_0\sim \rho/2$,
$\pi_{\e,\pm\mp}/\pi_{\e,\pm\pm}\sim 20$.  That is, the lens will
transit the source at similar (source) impact parameters, and it
would require fine tuning to arrange that they transited at
almost symmetric impact parameters\footnote{\ch We note that for
the very large sources, $\rho\gg 1$, parallax measurements may be difficult
for the subset of events with $z_0\ll 1$, where $z\equiv u/\rho$.
The ``effective half-duration'' $t_{\rm dur} \equiv t_*\sqrt{1-z_0^2}$ is
extremely well determined from the light curve, but a small fractional
error in $t_*$, $\delta\ln t_*$ then results in comparable error in $z_0^2$,
$\delta(z_0^2)\simeq 2\delta\ln t_*$, and hence (for $z_0\ll 1$), a much
larger error in $z_0$, and so in $u_0$: 
$\delta u_0\simeq \rho^2\delta t_*/u_0$
See, e.g., the case of OGLE-2012-BLG-1323 \citep{ob121323}.
However, these concerns do not apply for larger $z_0$, such as the
$z_0=0.5$ example used here to illustrate the applicability of
the Rich Argument to large sources.}.  
{\ch In addition, for} cases that
$\rho\ll 1$, so that $u_0\ll \rho$ is necessarily small as seen
from one observatory (to have an FSPL event), then the lens trajectory
may fall well outside the source as seen from the other observatory,
in which case $\pi_{\e,\pm,\pm} \simeq \pi_{\e,\pm,\mp}$, so there is no
real degeneracy.  However, particularly for $\rho\sim 1$, which
includes the most extreme and difficult lenses, there may be significant
ambiguity in the interpretation of individual events.

This discrete degeneracy interacts with the continuous degeneracy
in $\Delta u_0$.  If the source flux is left as a free parameter
for a 1L1S (more specifically FSPL) event,
then the error in $u_0$ will in general be much larger than the
error in $t_0/t_\e$.  Therefore, if the two $u_0$ from the two 
observatories are treated as independent parameters, then the error
in $\Delta u_0$ will be correspondingly greater than in $\Delta t_0/t_\e$.
However, if the ratio of source-flux parameters $f_s$ 
is constrained from comparison stars, then
the error in $\Delta u_0$ can be greatly reduced, but only for the
small $\pi_\e$ solution.  See Equation~(2.5) of \citet{gould95} for the
first example of a calculation of this effect.  The reason is that
as the flux is varied, the two values of $|u_0|$ move in tandem.
For the small-parallax solution, this means that the two values of $u_0$
also move together, but for the large-parallax solution, they move oppositely.

For relatively bright sources, the issue of continuous degeneracies
can be removed if a good argument can be made that the source is unblended,
so that the source flux can be fixed. In the general case, the same
argument can ultimately be made after followup AO observations show
the source and (possible) host in isolation.  Then the source flux can
be measured (and transformed to $I$ band), thereby greatly reducing
the continuous degeneracy.  If the source and host
both appear, then the vector proper motion can be measured, which
will very likely completely resolve the four-fold degeneracy (see
Figure~1 of \citealt{gould94b}).  Of course, this will not include
the cases of genuine FFPs.  For some fraction of these, there will
be $\sim {\cal O}(2)$ errors in the mass estimate due to unresolved
four-fold degeneracies.

{\subsection{{Bound Planets: Masses and Distances}
\label{sec:bound}}

KMT+L2 would have many other applications.  We focus here on
those that rest on combined mass and distance determinations for
microlensing events, i.e., combined $\theta_\e$ and $\pi_\e$ measurements.
First among these are binary-lens single-source (2L1S) events, particularly
those containing planets.

\citet{graff02} pointed out that 2L1S events were ripe for mass measurements
by a parallax satellite because (in contrast to the overwhelming majority
of 1L1S events) ground-based data routinely return measurements of
$\theta_\e$ due to finite-source effects during caustic crossings.
This in turn is due to the facts that the caustics are much larger
for 2L1S and (very importantly) that we mainly become aware of the
lens binarity due to such caustic crossings.  Hence, all that is needed
to complete the mass and distance determinations is a $\pi_\e$ measurement.
\citet{graff02} investigated a number of problems related to such
measurements (including the role of degenerate solutions -- see their
Figure~4), but they did so within the context of an Earth-trailing
parallax satellite that would be triggered to sparse observations
by a ground-based alert.  

\citet{ggh} studied a problem much closer to the present one: a survey
telescope at L2, complemented by a simultaneous ground-based survey.
However, their main concern was to investigate the possibility of
measuring $\pi_\e$ and $\theta_\e$ for Earth-mass planets
even in the absence of a caustic crossing (see their Figure~1).  They
comment in passing that such a ground+L2 survey will routinely
yield $\pi_\e$ plus $\theta_\e$ measurements for caustic crossing
events, but they do not further explain this.

Here, we discuss to what extent this is actually the case.  We begin
by asking what can be learned from observations of the source
crossing a single caustic, combined with a model of the caustic
geometry derived from the overall light curve as observed from
a single observatory (say, the satellite).  In particular, the
crossing will take place at an angle $\phi$ (where $\phi=0$
corresponds to perpendicular).  Then, as seen from Earth,
the crossing will take place 
$\Delta t=\Delta t_0 + \tan\phi\Delta u_0 t_\e$ later as seen
from Earth, where $(\Delta t_0/t_\e,\Delta u_0)\tilde r_\e$ is the offset 
between the two observatories in the Einstein ring.  If
only $\Delta t$ is measured (no matter how precisely), one can
gain only one constraint on the vector 
$\Delta {\bf u}=(\Delta \tau,\Delta \beta)\equiv(\Delta t_0/t_\e,\Delta u_0)$,
and hence cannot uniquely determine $\pi_\e = \Delta u\,\au/D_\perp$.
In principle, there is additional information from
the difference between the ``strength'' of the caustic at the positions
crossed by the source as seen from the two observatories.  A ``stronger''
caustic will lead to a greater magnification at the peak of the
crossing.  However, except near the cusps, the gradient in caustic
strength is very weak, meaning that in practice it is difficult
or impossible to extract useful information from the different caustic-peak 
fluxes.  As pointed out by \citet{ggh}, it is also possible
to get an independent constraint from the one-dimensional annual parallax
\citep{gmb94} measured from the overall light curve.  This will be
feasible in some cases, but not others, in particular those with short
$t_\e$ and/or faint sources.  Here we focus on extraction of $\pi_\e$
from the caustic features of the light curve alone.

As caustics are closed curves, every entrance will be matched by an exit.
If delays $\Delta t_1$ and $\Delta t_2$ occur
at crossing angles $\phi_1$ and $\phi_2$, and each is measured with 
precision $\sigma$, then the measurements can be expressed as two
equations with two unknowns $(\Delta\tau,\Delta\beta)$,
\begin{equation}
\Delta t_i=(1\cdot\Delta\tau + \tan\phi_i\cdot\Delta\beta) t_\e \pm \sigma;
\qquad
(i = 1,2),
\label{eqn:dtaudbeta}
\end{equation}
whose covariance matrix is
\begin{equation}
c_{i,j}\biggl(\matrix{\Delta\tau \cr \Delta\beta}\biggr) =
{\sigma^2\over 
(\tan\phi_2-\tan\phi_1)t_\e^2}\biggl(\matrix{
\tan\phi_2 & -\tan\phi_1 \cr -1 & 1}\biggl) .
\label{eqn:covmat}
\end{equation}
Thus, for example, if the crossings are on consecutive caustic segments
(i.e., separated by one cusp) then $|\phi_2 - \phi_1|$ is likely to
be a few tens of degrees, so that $|\tan\phi_2 - \tan\phi_1|^{-1}$ will
be only of order a few.  However, if the caustic segments are separated
by two cusps (or three cusps for some resonant caustics), then they
could be roughly parallel, leading to $|\tan\phi_2 - \tan\phi_1|^{-1}$ being
of order ten or even a few tens.

However, in most cases, the measurement of two different caustic-crossing
time offsets will yield good mass and distance determinations.
The first point is that the offsets themselves are of order
\begin{equation}
\Delta t \sim {D_\perp\over \tilde v} = {D_\perp\over \au}\,{\pi_\rel\over\mu_\rel}
= 13\,{\rm min}{D_\perp\over 0.01\, \au}
\biggl({\pi_\rel\over 16\,\muas}\biggr)
\biggl({\mu_\rel\over 6.5\,\masyr}\biggr)^{-1},
\label{eqn:deltat}
\end{equation}
while the full caustic crossing times will be 
$2\,|\sec\phi|t_* = 97\,{\rm min}\,\sec\phi(\theta_*/0.6\,\muas)/
(\mu_\rel/6.5\,\masyr)$.  Hence, there will be many observations per crossing.
Here, $\tilde v \equiv \tilde r_\e/t_\e$ is the lens-source relative
velocity projected on the observer plane.
Moreover, caustic crossings generally yield magnification jumps 
$\Delta A\sim {\cal O}(10)$, which are easier to detect than the
$\Delta A\sim$few, level events that define the requirements of the
FFP experiment.

The main difficulty is that both observatories must observe both caustics.
This would be automatic for the L2 satellite during the time (perhaps eight
months per year) of its continuous observation.  However, the ground
observations would face interruptions due to weather and diurnal cycle.
Approximating the two caustic crossings as independent, and scaling
weather/Moon interruptions as (15,25,35)\% at KMT(C,S,A),
then 22\% of all events during the (365 day) year would be observed
during two caustic crossing and another 29\% would be observed during 
one.  See Figure~\ref{fig:obs}. 
As noted above, $\pi_\e$ could be recovered for 
some of the latter by combining the single caustic time offset with the
1-D annual parallax signal.  Moreover, high-magnification events,
which are especially prone to planetary anomalies \citep{griest98},
will often yield parallax measurements even if planetary caustic
crossings are not observed \citep{yee13}.
We also note that if we exclude the 31 days
closest to opposition, when both parallax signals will be much smaller
(due to small $D_\perp$ and low projected acceleration of the satellite
and of Earth), then these percentages drop to 17\% and 26\%, respectively.

{\section{{Two-Satellite Experiment: IRx2}
\label{sec:ideal}}

By launching two identical survey telescopes into orbits separated
by $D\sim {\cal O}(0.01\,\au)$, one could pursue substantially different
science relative to KMT+L2 (Section~\ref{sec:kmt+l2}).  For example,
one satellite could be at L2 and the other in a low-Earth polar orbit or
in geosynchronous orbit at relatively high inclination\footnote{\ch
Another, more ambitious, approach would be to launch three such satellites
into the same L2 halo orbit with epicyclic radius, e.g.,
$r_{\rm halo} \sim 0.003\,\au$, and separated in phase by $120^\circ$.
Then, the projected separation between some pair of these would
always be $1.5\,r_{\rm halo}< \max_{ij}(D_{\perp,ij}) <\sqrt{3}\,r_{\rm halo}$,
where $D_{\perp,ij}$ is the projected separation between satellites $i$ and $j$.
Then the third satellite could almost always break the 
$\pi_{\pm\pm}/\pi_{\pm\mp}$ degeneracy, even when $\min_{ij}(D_{\perp,ij})$ 
was small
(although the much less important directional degeneracy would usually
not be broken).  \citet{bachelet19} and \citet{ban20} discuss a 2-satellite
variant of this scenario, composed of the planned {\it Euclid} and
{\it Roman} missions, although the possibility of joint observations
is restricted to about 40 days per year by design features of these
observatories.  See Section~\ref{sec:par-only}.
}.

The main value of having two identical satellites is that the experiment
would not be fundamentally constrained by the limits of ground-based
observations. These constraints include both time coverage and resolution.  
But the most important ground constraint comes from the high sky
background in the infrared (IR).

To address the various choices, we first focus on the main potential 
scientific objectives.  As we have discussed above, the effective
limit of KMT+L2 in Einstein radii is $\theta_\e\ga 1.5\,\muas$,
which roughly corresponds to FFP masses $M\sim 6\,M_\oplus$ in the bulge
or $M\sim 0.7\,M_\oplus$ in the disk.  Bodies of the latter mass scale
are relatively common in the solar system (two examples).  However, bodies that
are 100 times less massive, i.e., $M\sim 0.5 M_{\rm Moon}$ are more
common, even though they are substantially more difficult to detect.
In the context of the Solar System, 
such bodies could plausibly have been prodigiously ``ejected'' to 
the Oort Cloud or to unbound orbits, or they could remain ``hidden''
in the outer regions of the Kuiper Belt.   Hence, the systematic
study of such objects, both bound to and unbound from other stars,
would give enormous insight into planetary-system formation and
evolution.  In particular, the FFP candidates with measured parallaxes
and vector proper motions could be identified as part of the Oort
Cloud of their hosts, even at several $10^4\,\au$, because the 
``background'' of ambient field stars could be drastically reduced
by demanding common proper motion and distance \citep{gould16}.

To reach the goal of $\theta_\e \sim 0.15\,\muas$ requires one
to probe $\theta_*\sim 0.3\,\muas$ sources, which
corresponds to $M_{\rm source}\sim 0.5\,M_\odot$.  In this case,
$\rho \sim 2$, so $A_\max \simeq \sqrt{1 + 4/\rho^2}\sim 1.4$.
These $M_I\sim 6.5$ sources are a few times more common than G dwarfs,
but also have only half the cross section $(2\,\theta_*)$.  Therefore,
the underlying rates are similar.  The main difficulty is that the 
difference-star magnitude at $A_\max = 1.4$ and extinction $A_I=2$
is $I_{\rm diff} = 24.1$.  To obtain 10\% photometry in a 9-minute
diffraction-limited exposure would require a $D\sim 2$m diameter mirror.

An alternative would be a broad $H$-band filter similar to that of the
{\it Nancy Grace Roman} (f.k.a {\it WFIRST}) telescope \citep{spergel13}. 
At $H_{\rm diff} = 21.1$, a 9-minute exposure on a 0.5m telescope
(with diffraction-limited FWHM$\sim 0.8^{\prime\prime}$), would yield 10\% 
photometry.  The same camera layout as KMT+L2 ($0.4^{\prime\prime}$ pixel scale,
18k$\times$18k detectors) would then imply Nyquist sampling.  Thus,
the telescope dimensions are qualitatively similar to KMT+L2,
the main difference being that the former would be equipped with IR detectors.
We dub this two-telescope system: IRx2.

Simultaneous observation by two identical telescopes plays a central
role not only in measuring $\pi_\e$ (and so the FFP masses), but also
in robustly distinguishing between extremely short, very faint microlensing
events and various forms of astrophysical and instrumental noise.

The key problem is that for each square degree, and for a year of
integrated observations. IRx2 would observe
about $3\times 10^7$ early M dwarfs for 8760 hours, enabling
$2.5\times 10^{11}$ independent probes for 
dwarf-planet FFPs/Exo-KBOs/Exo-OCOs.  This might yield five detections
or 5000 (per yr-deg$^2$): there is no way to reliably estimate at this
point.  But how can one be sure that any one, or any 1000 of these
are actually due to microlensing?  One issue is instrumental noise.
If the noise were Gaussian, then an $8\,\sigma$ signal would be enough
to reject false positives at $p=\exp(-8^2/2)\sim 10^{-14}$.  However, it would
be difficult to rule out non-Gaussian noise in the detector or the
detection system based on one observatory alone (such as
{\it Roman}, which is expected to detect of hundreds of larger FFPs, 
\citealt{johnson20}).  But this possibility
would easily be ruled out if two observatories saw an event on the same
star at nearly the same time.

A more fundamental problem is astrophysical noise.  Suppose that
one in a million early M dwarfs had one one-hour outburst per year.
This would give rise to 30 ``FFP events''.  In fact, M dwarfs are known
to have outbursts\footnote{Of course, just as with cataclysmic
variables in current microlensing surveys, one would have to begin
by eliminating all light curves with more than one outburst over
the lifetime of the experiment.  Only then would it be feasible to
vet the relatively few remaining ``bumps'' by comparing the observations
of the two satellites.}.  In contrast to the case of instrumental noise,
merely observing the same event from two independent telescopes would
not guard against this astrophysical noise in any way: the effect is
real, so all observers viewing it from the same place will see the same
thing.

However, for IRx2, the event will look different as seen from
the two observatories, i.e., delayed and/or with a different amplitude.
The delay (tens of minutes) will be much longer than the light travel 
time between the observatories ($<5\,{\rm sec}(D/0.01\,\au)$).  There
is no other effect that could cause delays, apart from interstellar
refraction, which is not a strong effect at these wavelengths.  Similar
arguments apply to differences in the amplitude of the event as seen
by the two observatories.

{\section{{Parallax-only Events}
\label{sec:par-only}}

Our focus in this paper has been on FFP mass measurements, 
$M=\theta_\e/\kappa\pi_\e$, derived from simultaneous measurements
of $\theta_\e$ and $\pi_\e$.  However, any experiment capable of
delivering both parameters will yield many more measurements of
$\pi_\e$ that are not complemented by measurements $\theta_\e$.
In this section, we investigate the relative precision of mass
{\it estimates} based on $\pi_\e$-only measurements, compared to
{\it measurements} based on $\pi_\e$+$\theta_\e$.  We then ask
in what way and to what degree such estimates can augment our
understanding of the FFP population and of other low-mass, non-luminous
objects.

There are several previous studies that focused on L2-scale
microlens parallax measurements.  \citet{zhu16} investigated combined
KMTNet and {\it Roman} observations under the assumption that the latter
would point at a relatively unextincted $(A_H\sim 0.5)$ field.
In contrast to the KMT+L2 study that we carried out in 
Section~\ref{sec:kmt+l2}, the \citet{zhu16} L2 telescope was vastly
more powerful than KMT, so that any event that was detectable
by KMT had essentially perfect data from {\it Roman}.  Nevertheless,
we can use this study for some basic guidance to the issues discussed 
below.  

\citet{bachelet19} and \citet{ban20} each studied joint observations by
{\it Euclid} and {\it Roman}, both of which are planned to have L2 orbits.
As mentioned above, ``L2 orbits'' have epicyclic
radii of few$\times 10^5\,$km, around
the mathematical ``L2'' point, so that the separation of these two
satellites could be a large fraction of the Earth-L2 distance.  They
pursued complementary approaches. \citet{bachelet19} applied a Fisher-matrix
analysis to a narrow subset of possible events, while \citet{ban20}
subjected a detailed Galactic-model simulation to relatively simple cuts.
In addition, \citet{ban20} investigated joint LSST-{\it Roman} observations,
as well as some other combinations.

All three studies note that finite-source effects impact 
an increasing fraction of events as the lens mass decreases.   However,
only \citet{zhu16} make a quantitative estimate of this fraction. 
See their Figure~2.

{\subsection{{Precision of Parallax-only Mass Estimates}
\label{sec:precision}}

Although satellite microlens-parallax measurements were proposed
more than a half century ago \citep{refsdal66} and have been
a very active area of investigation for a quarter century \citep{gould94b},
it appears that no one has asked what is the fundamental limit
of parallax-only microlens mass estimates.  The key to doing so is a point
already made by \citet{han95}: lens populations at different distances
have very similar proper-motion distributions, $\mu_\rel=\theta_\e/t_\e$, 
but very different projected velocity distributions, 
$\tilde v = (\au/\pi_\rel)\mu_\rel$.  Therefore, by measuring $\theta_\e$,
or (because $t_\e$ is usually very well measured) equivalently $\mu_\rel$,
one is only determining more precisely a quantity that is 
already ``basically known''.  By contrast, once $\pi_\e$ is measured, one
can ``guess'' $\mu_\rel$ to reasonable precision and then estimate
$M=\mu_{\rel,\rm est}t_\e/\kappa\pi_\e$.  Indeed this is how \citet{ban20}
estimated lens masses from the simulation results, which were then compared to
the simulation input masses.  \citet{ban20} found, using 
$\mu_{\rel,\rm est}=7.5\pm 1.5\,\masyr$, that the scatter was larger than
the assumed error, although the exact origin
of this discrepancy could not be pinpointed because the simulation contained 
several other sources of error.

Let us initially consider a lens that is ``known'' to be in the bulge.  
For example, it could have $\tilde v=3000\,\kms$.  Let us assume
that $\pi_\e$ and $t_\e$ are measured very well.  We can imagine
trying to evaluate the lens mass by a Bayesian analysis
using a Galactic model.  If we ignore for the
moment any prior on the lens mass, then the only relevant information
from the model is the kinematic distributions of the sources and
lenses, which are both drawn from the same approximately isotropic
Gaussian, with dispersion $\sigma$.  Then, this Bayesian analysis will
return an analytic result (see Appendix),
\begin{equation}
M_{\rm est} = {\mu_{\rel,\rm est} t_\e\over\kappa\pi_\e}
={t_\e\over\kappa\pi_\e}[\langle\mu_\rel\rangle \pm \sqrt{{\rm var}(\mu_\rel)}]
={\sigma t_\e\over\kappa\pi_\e}\biggl[{4\over\sqrt{\pi}}\pm \sqrt{6-{16\over\pi}}
\biggr],
\label{eqn:massest}
\end{equation}
which then yields a fractional error in the mass estimate
\begin{equation}
{\sqrt{{\rm var}(M)}\over M_{\rm est}} = 
{\sqrt{{\rm var}(\mu_\rel)}\over \mu_{\rel,\rm est}} 
= \sqrt{{3\pi\over 8} -1} = 0.42 .
\label{eqn:massest2}
\end{equation}
If we now add a mass prior, then it could somewhat change the mean
estimate, but (unless it is very strong), it will not substantially change
the standard deviation.  Note that in the regime of FFPs, there is
no basis for a strong prior (otherwise we would not be doing the
experiment).

Because the $\mu_\rel$ distribution for disk lenses is very similar to that 
of bulge lenses, the fractional mass error is likewise similar.  We have
not as yet included the directional information that is returned by the
$\bpi_\e$ measurement.  However, for bulge lenses, this is completely
irrelevant because the prior is essentially isotropic.  And it is basically
irrelevant if the lens is known to be in the disk because the directional
distribution for $\bmu_\rel$ is basically independent of distance.
For cases that the lens could be either in the disk or bulge, the
directional information can help distinguish between these alternatives,
but this cannot reduce the fractional mass-estimate error below
the values derived under the assumption that the population is known.

Therefore, there is a hard limit of $42\%$ fractional mass-estimate
error, even if the parallax is measured perfectly\footnote{Strictly
speaking, this statement applies only if there is literally no information
about $\rho$.  However, even if $\rho$ is not measured, there could
in principle be an upper limit on $\rho$. The actual upper limit must
be determined from the fit, but in general it is of order $\rho\la u_0$.
Then $\mu_\rel \ga \theta_*/u_0 t_\e$.  Indeed, \citet{kb192073}
used this formalism to identify FSPL candidates.  For the general case,
the fraction of $\pi_\e$-only events that have such partial $\rho$
information will be small, but it will be larger for the lowest-mass
FFPs because the lens must then pass within a few $\theta_*$ to yield
a parallax measurement.  See Section~\ref{sec:role}.}.

{\subsection{{Improvement from Measuring $\bmu_s$}
\label{sec:improve}}

For the case of FFPs, there is no lens light, so the source proper motion
$\bmu_s$ can be measured from two well separated high-resolution
epochs.  For example, we have suggested {\it CSST} observations.

Let us first imagine that such a measurement has been made, and it is
found that $\bmu_s = 0$ (in the bulge frame).  Then, given this measurement,
the estimated $\bmu_\rel$ distribution still has a mean of zero, but
now its dispersion drops from $\sqrt{2}\sigma$ to $\sigma$.  This
means that the mass estimate drops by a factor $\sqrt{2}$ from what
it would be without this measurement, but the fractional mass error,
given by Equation~(\ref{eqn:massest2}) is still exactly the same.
However, as $|\bmu_s|$ increases (in the bulge frame), there is increasing
information constraining both the magnitude and direction.  To illustrate
this, we initially ignore the directional information and show the
resulting mean mass estimate and fractional mass-estimate error (relative
to the case of no source proper-motion measurement) in
Figure~\ref{fig:muave}.

The role of directional information is difficult to represent, in
part because there are two directions that cannot be distinguished
for short events without a third observatory at a similar distance
from the first two (and even with such an observatory, breaking
this directional degeneracy is difficult, \citealt{mb16290}).  Therefore, we
do not further pursue this analytic approach.  The importance of the
directional information can only be assessed by extensive Monte Carlo
simulations.  Nevertheless, as we have shown in Figure~\ref{fig:muave},
$\bmu_s$ measurements can definitely contribute to the understanding
of $\pi_\e$-only events.

{\subsection{{Role of Parallax-Only Measurements}
\label{sec:role}}

In the KMT+L2 experiment that we have described, for
every bulge super-Earth with measurements of both $\theta_\e$ and
$\pi_\e$, there will be of order one with a $\pi_\e$-only measurement.
This can be understood by considering Figures~\ref{fig:lc0} and 
\ref{fig:lcerr}: if the lens had passed outside the source
(so dropping in amplified flux by a factor $\sim 2$) then the event
would have already become much noisier.  An additional factor of two
would make parameter measurements difficult or impossible.  \citet{zhu16}
reached similar conclusions for somewhat different assumptions.
For the $t_\e\sim 0.3\,$day bulge events that we are considering,
about half of all those with $\pi_\e$ measurements also had $\theta_\e$
measurements.  See their Figure~2.  Because the $\pi_\e$-only mass
estimates have much larger errors, the addition of a comparable number
of these to the $\pi_\e$+$\theta_\e$ sample will have marginal
scientific impact.

However, the same is not true for higher-mass objects, such as free-floating
Jupiters and low-mass brown dwarfs.  First, these are almost certainly
much rarer, and the number of FSPL events is directly proportional
to the number density of the population.  Hence, the number of direct
mass measurements is likely to be tiny.  Second, the ratio of 
$\pi_\e$-only to $\pi_\e$+$\theta_\e$ measurements is much greater.
Consider for example, a lens that is 100 times more massive than
the $\theta_\e\sim 1.5\,\muas$ example in Figures~\ref{fig:lc0} and 
\ref{fig:lcerr}, i.e., right in the middle of the ``Einstein desert''
discussed by \citet{kb172820}.  With the same trajectory, it would
be about 10 times brighter.  Even passing at $\sim 5\theta_*$, it would
be somewhat brighter than the event in those figures.  Thus, $\pi_\e$-only
measurements may be the only way to obtain mass estimates for this
population.

{\section{{Discussion}
\label{sec:discuss}}

The basic physical principle, i.e., synoptic
observations of FSPL events from two locations, is the same as
that proposed by \citet{gould97} for terrestrial-parallax mass 
measurements of ``extreme microlensing'' events.  \citet{gould13}
later showed that the expected rate of such ``extreme microlensing''
mass measurements was only of order one per century under observing
conditions of that time.  Hence, they found it surprising that there
were already two such measurements \citep{ob07224,ob08279} when they
published their analysis..

Because the physical principle and its mathematical representation are
identical, it is worthwhile to understand the physical basis of the
${\cal O}(10^{3.5})$ difference in expected rates relative to the 
KMT+L2 experiment that we describe here.

Several factors are actually similar, including
$\langle \mu \rangle = (10\,{\rm vs.}\,6.5)\masyr$, and
$\langle \theta_* \rangle = (0.6\,{\rm vs.}\,0.5)\muas$, as
well as the assumption that $(25\,{\rm vs.}\,28)\%$ of the
year would be effectively monitored.  Moreover, the total number
of sources assumed by \citet{gould13} was about 10 times higher
because they considered potential follow-up observations of
all Galactic bulge microlensing fields, whereas we have assumed
continuous observations of just $4\,{\rm deg}^2$.  However,
this enhancement was canceled by the fact that only 1/10 of
events would be observable at peak from multiple continents.
Moreover, they estimated that only half of these would be
successfully monitored (based on the statistical analysis of
\citealt{gould10}).

The overwhelming majority of the difference comes from the fact that
\citet{gould13} estimated a surface density of lenses of 
$4.5\times 10^2\,{\rm arcmin}^{-2}$, whereas we have estimated
$5\times 10^5\,{\rm arcmin}^{-2}$, i.e., a difference of $10^3$.
A small part of this difference was in turn due to a different FFP
model.  In both cases, the lens surface density is dominated by
FFPs, but \citet{gould13} used the \citet{sumi11} model of two FFPs
per star, whereas we have used the \citet{mroz17} model of five FFPs
per star.  However, the main difference was that \citet{gould13} showed
that the FSPL+terrestrial-parallax technique could only be applied
to lenses within $D_L\la 2.5\,\kpc$.  That is, even with very good photometry
on highly magnified\footnote{Note that at $D_L=2.5\,\kpc$, even a 
\citet{sumi11} $M=M_j$ planet has $\theta_\e\sim 50\,\muas$ and so
$A_\max\sim 170$.  This compares to $A_\max\sim 5$ for a \citet{mroz17}
$M=5\,M_\oplus$ planet in the Galactic bulge, $\pi_\rel\sim 16\,\muas$.}
events (such as OGLE-2007-BLG-224 and OGLE-2008-BLG-279)
they considered that only events with $D_\perp/\tilde R_*>0.02$ would
yield measurable parallaxes.  Because $D_\perp\sim R_\oplus$ for
terrestrial parallax (compared to $D_\perp\sim 250\,R_\oplus$ for L2 parallax),
and because $\tilde R_* = \au\theta_*/\pi_\rel$, the FSPL+terrestrial-parallax
technique is restricted to lenses with high $\pi_\rel$.

{\section{{Conclusions}
\label{sec:conclude}}

The masses of FFPs can only be measured over a broad range
of distances by simultaneous microlensing
surveys conducted by two observatories separated by ${\cal O}(0.01\,\au)$.
Fortuitously, KMTNet can operate as one of these observatories.  We show
that a 0.3m telescope at L2, equipped with a KMT-like camera could be 
the other.  Such a system would measure the masses of about 130 FFPs 
(from the known super-Earth population) over
a 4-year mission, taking account of breaks in the observing schedule
due to weather, Moon, and the diurnal and annual cycles.  It could
also discover lower-mass FFPs in the disk (down to Earth-mass or below), 
if these are equally common or more common.  Finally, it would
measure the masses and distances of many bound planets.

A next generation experiment, IRx2, consisting of two 0.5m IR 
satellites (at L2 and near Earth), could probe to sub-Moon masses,
which are generally classified as ``dwarf planets'' rather than
``planets''.  These might actually be ``free'', but could also
be exo-KBOs and exo-OCOs.  Once the 
mass, distance, and proper motion of these objects is found
by IRx2, one epoch of AO followup can distinguish between these objects
being ``free'' or part of exo-systems.  If the latter, a second AO epoch
can measure their projected separation from their host, and so determine
whether they are exo-KBOs or exo-OCOs.

The duality and separation of the IRx2 system is crucial for verifying
that the very weak and rare signals due to dwarf-planet FFPs are
caused by microlensing rather than instrumental or astrophysical effects.
The fact that there is signal from both satellites will prove that
it is not due to instrumental effects.  And the fact that the two
signals are different (due to parallax) will prove that they are not
due to astrophysical effects.

\acknowledgements

We thank Jennifer Yee for insightful comments on the manuscript.
W.Z. and S.M. acknowledges support by the National Science Foundation
of China (Grant No. 11821303 and 11761131004). S.D. is supported by
National Key R\&D Program of China No. 2019YFA0405100.

\appendix
\section{$\langle\mu\rangle$ for Isotropic Gaussian}
\label{sec:append}

If $\bmu_l$ and $\bmu_s$ have isotropic 2-D Gaussian distributions
with the same dispersions $\sigma$ and the same mean, 
then $\bmu_\rel =\bmu_l -\bmu_s$ 
has a Gaussian distribution with zero mean and dispersion $s=\sqrt{2}\sigma$:
\begin{equation}
f(\mu_x,\mu_y)d\mu_x\, d\mu_y = {\exp[-(\mu_x^2 + \mu_y^2)/2s^2]
\over 2\pi s^2}d\mu_x\, d\mu_y
              = {\exp[-\mu^2/2s^2]\over 2\pi s^2}d\mu_x\, d\mu_y,
\label{eqn:a1}
\end{equation}
where for simplicity, we have relabeled $\mu = \mu_{\rel}$.
Then the mean value of $\mu$ weighted by the event rate (i.e., $\mu$ itself) is
\begin{equation}
\langle\mu\rangle = {\int d\mu_x\,d\mu_y\, \mu\times \mu \,f(\mu_x,\mu_y)
\over\int d\mu_x\,d\mu_y\, \mu\, f(\mu_x,\mu_y)}
= {\int_0^\infty2\pi\mu\,d\mu \mu^2 \exp (-\mu^2/2s^2) \over
\int_0^\infty2\pi\mu\,d\mu \mu \exp (-\mu^2/2s^2) }
\label{eqn:a2}
\end{equation}
\begin{equation}
\langle\mu\rangle = 2\,\sigma{\int_0^\infty dz\,z\exp(-z)\over
\int_0^\infty dz\,z^{1/2}\exp(-z)}
= 2\,\sigma{1!\over (1/2)!} = {4\over\sqrt{\pi}}\sigma,
\label{eqn:a3}
\end{equation}
where $z\equiv \mu^2/4\sigma^2$.  Similarly,
\begin{equation}
\sqrt{{\rm var}(\mu)}=\sqrt{\langle\mu^2\rangle -\langle\mu\rangle^2} =
\sqrt{6 -{16\over\pi}}\sigma.
\label{eqn:a4}
\end{equation}



%

\begin{deluxetable}{lcc}
\tablecolumns{3} \tablewidth{0pc} \tablecaption{\textsc{KMT+L2 Telescope Properties}} 
\tablehead{\colhead{Property} & 
           \colhead{KMT} &
           \colhead{L2}} \startdata
Aperture (m)        & 1.6 & 0.3 \\
Cycle time (min)     & 8.25 & 10.0 \\
$I$-band zero point & 29.20 & 26.75 \\
Pixel Size (arcsec) & 0.40 & 0.40 \\
FWHM (arcsec)       & 1.3--2.5 & 0.67 \\
$I$-background (arcsec$^{-2}$) & 18.5 & 20.5 \\
$I$-background (PSF$^{-1}$) & 16.7--15.3 & 20.1 \\
$V$-band zero point & 28.65 &  \\
$V$-background (arcsec$^{-2}$) & 20.9 & \\
\enddata
\tablecomments{KMT ``cycles'' contain three 60s $I$- and one 75s
$V$-band exposures.  KMT background is sky dominated and is taken
as 1.4 times ``dark'' value. L2 background is ambient-star dominated
and is evaluated at $A_I=2$ and $2\times$\citet{holtzman98}
luminosity function.  Compare baseline and magnified performance
of KMT vs.\ L2 in Figure\ref{fig:lc0}
}
\label{tab:telescopes}
\end{deluxetable}

\begin{deluxetable}{lccccc}
\tablecolumns{3} \tablewidth{0pc} \tablecaption{\textsc{FFP Mass Measurements for KMT+L2}} 
\tablehead{\colhead{Characteristic} & 
           \colhead{G dwarfs} &
           \colhead{MS/SG} & 
           \colhead{Lower Giants} & 
           \colhead{Upper Giants} & Total } \startdata
$\langle \theta_* \rangle\ (\muas)$    &0.5  & 1.2 & 4.5 & 7.0 & \\
$N_{\rm source}\ ({\rm arcmin}^{-2})$ &3000  &800 &80 & 75 & \\
$2\langle\theta_*\rangle\langle\mu\rangle N_{\rm FFP,bulge} N_{\rm source}\Omega 
T_{\eff,\rm bulge}$  &44  & 28 & 10 & 15 & 97 \\
$2\langle\theta_*\rangle\langle\mu\rangle N_{\rm FFP,disk} N_{\rm source}
\Omega T_{\eff,\rm disk}$  &18  & 7 & 3 &4 & 32 \\
\enddata
\tablecomments{Assumes $\langle\mu\rangle = 6.5\,\masyr$, survey area
$\Omega= 4\,{\rm deg}^2$, $T_{\eff,\rm bulge}= 28\%(4\,{\rm yr})$,
$T_{\eff,\rm disk}= 37\%(4\,{\rm yr})$,
$N_{\rm FFP,bulge}=5\times 10^5\,{\rm arcmin}^{-2}$,
$N_{\rm FFP,disk}=N_{\rm FFP,bulge}/5$, and that there are 1.4 times more
accessible main-sequence (``G dwarf'') source for disk FFPs compared
to bulge FFPs.
}
\label{tab:sources}
\end{deluxetable}

\begin{figure}
\plotone{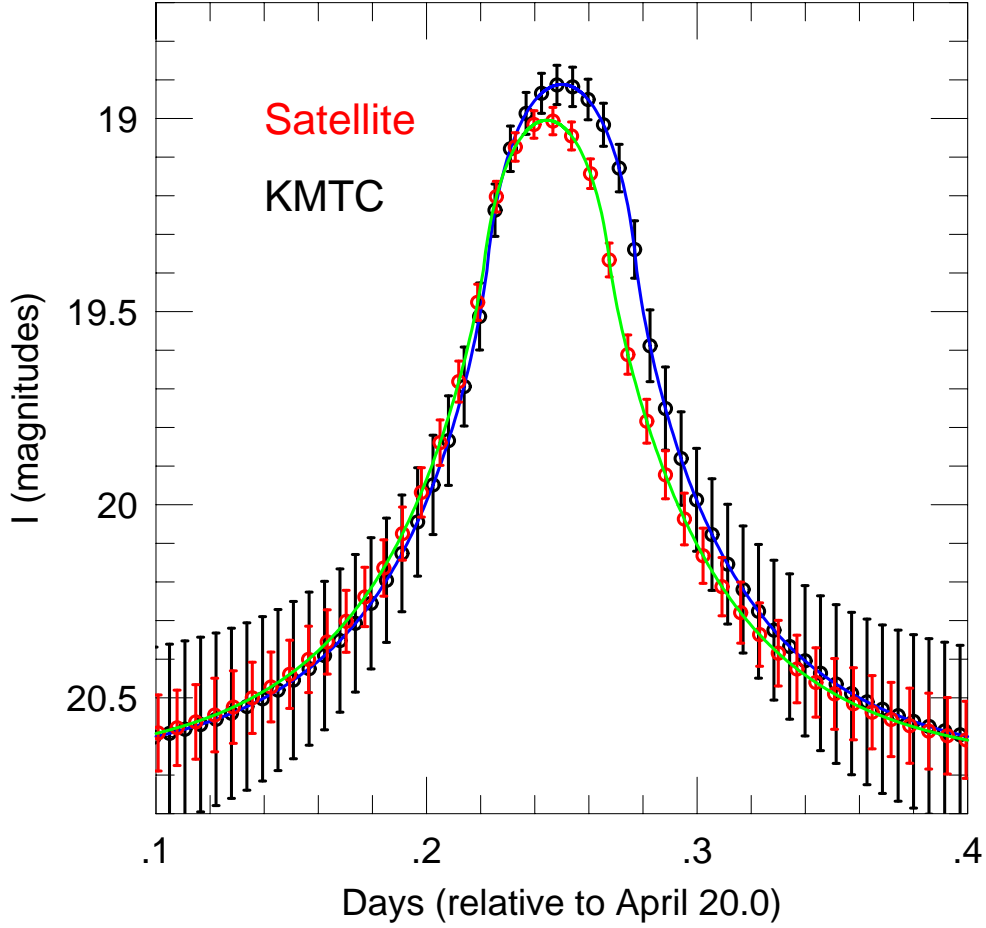}
\caption{Illustration of light-curve data and {\ch error bars} for KMTC (black)
and a 0.3m satellite such as Earth 2.0 (red) for an $M=5.75\,M_\oplus$
lens lying in the Galactic bulge, $\pi_\rel=16\,\muas$ (e.g., $D_S = 8.5\,\kpc$,
$D_L = 7.5\,\kpc$) magnifying a solar type source ($M_I=4$, $A_I=2$).
These parameters yield $\theta_*=0.58\,\muas$. $t_\e=120\,$min, $t_*=46\,$min,
$\rho=0.38$.  For Earth $(t_0,u_0) = (0.2500\,{\rm day},0.2000)$,
and for the satellite $\Delta(t_0,u_0) = (-7.9\,{\rm min},0.065)$.
The displayed KMTC points are binned over 8.25 minute cycles, of
three 1.00-minute $I$-band exposures, together with one 1.25 minute $V$-band
exposure (not shown), with 1.00 minute read-out time.  The satellite
points have 9.0-minute integrations and 1.0-minute read out.
In this case $\Delta t_0/t_* = 17\%$.  The four-fold degeneracy 
\citep{refsdal66,gould94b} yields two solutions with the true parameters
of the system, and another two with 
$(\Delta t_0,|\Delta u_0|) = (-7.9\,{\rm min},0.465)$, which would
imply $M=1.12\,M_\oplus$, $\pi_\rel= 82\,\muas$, and $D_L = 4.8\,\kpc$.
This solution requires fine tuning and would be heavily discounted
in a statistical analysis, but could not be ruled out in any individual
case.  See text.  {\ch For a concrete realization with simulated noise,
see Figure~\ref{fig:lcerr}.}
}
\label{fig:lc0}
\end{figure}

\begin{figure}
\plotone{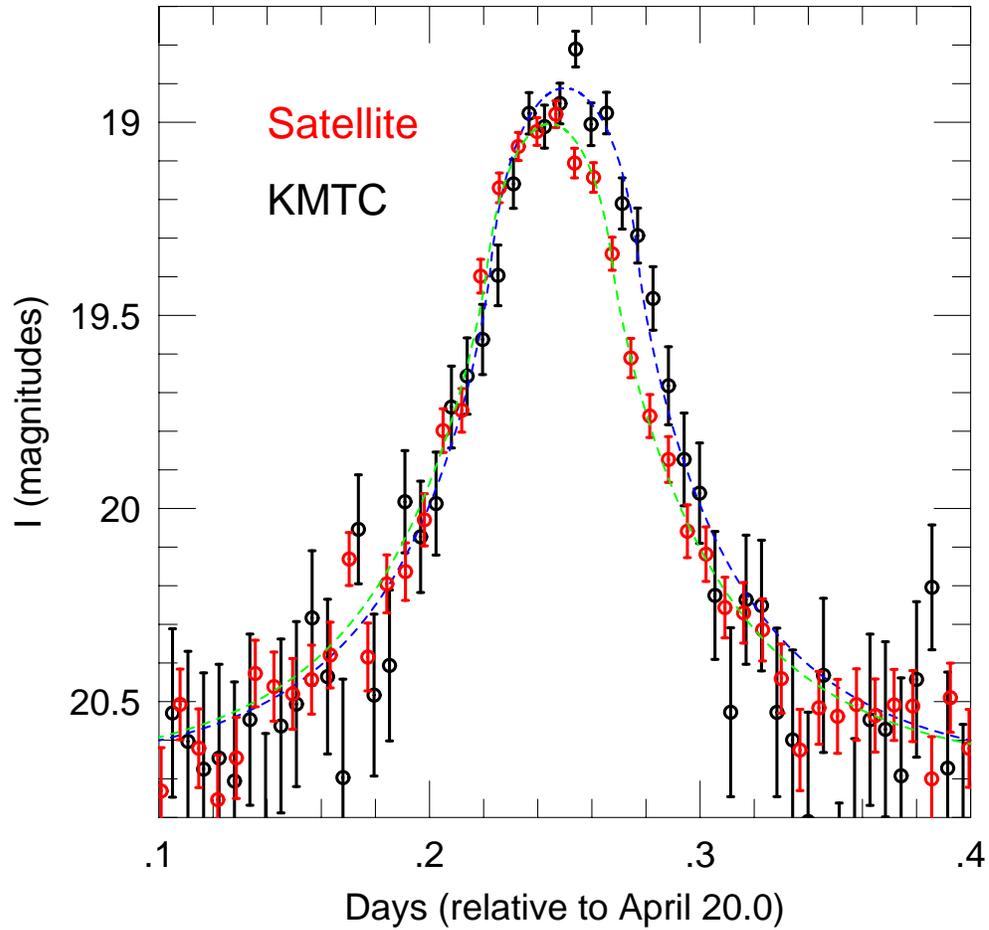}
\caption{\ch Concrete realization of observations of the event shown in 
Figure~\ref{fig:lc0} with simulated Gaussian noise.  The noise is applied
to the flux, with the values and errors transformed to magnitudes for
plotting.  Note that, even in the presence of scatter, the differences
in center, width, and height between the two curves is discernible by
eye.
}
\label{fig:lcerr}
\end{figure}

\begin{figure}
\plotone{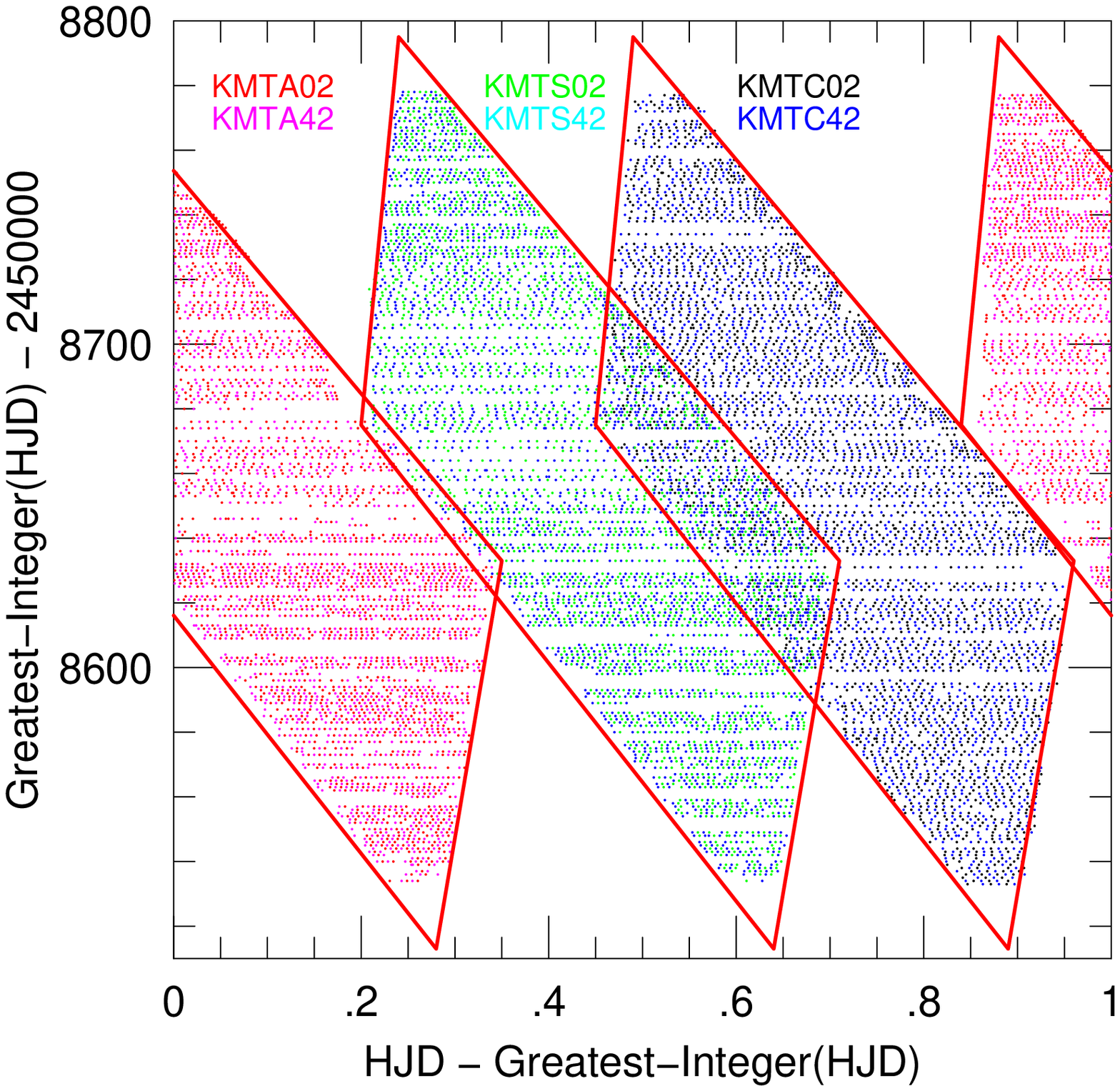}
\caption{Day versus time of actual 2019 observations
of closely overlapping KMT fields BLG02 and BLG42,
color coded as indicated in the legend for the three KMT observatories:
KMTA, KMTS, and KMTC.  The red quadrangle around the KMTC observations
is the empirically determined limit of the observational window
for that field.  Note that KMT does not observe in the extreme
wings of the season, but could in principle.  The remaining two
red quadrangles are translated versions of the KMTC quadrangle.
They match the empirical boundary of KMTS very well, but the match
is less perfect for KMTA.  Nevertheless, it is satisfactory for
present purposes. See Figure~\ref{fig:obs}.
}
\label{fig:kmt}
\end{figure}

\begin{figure}
\plotone{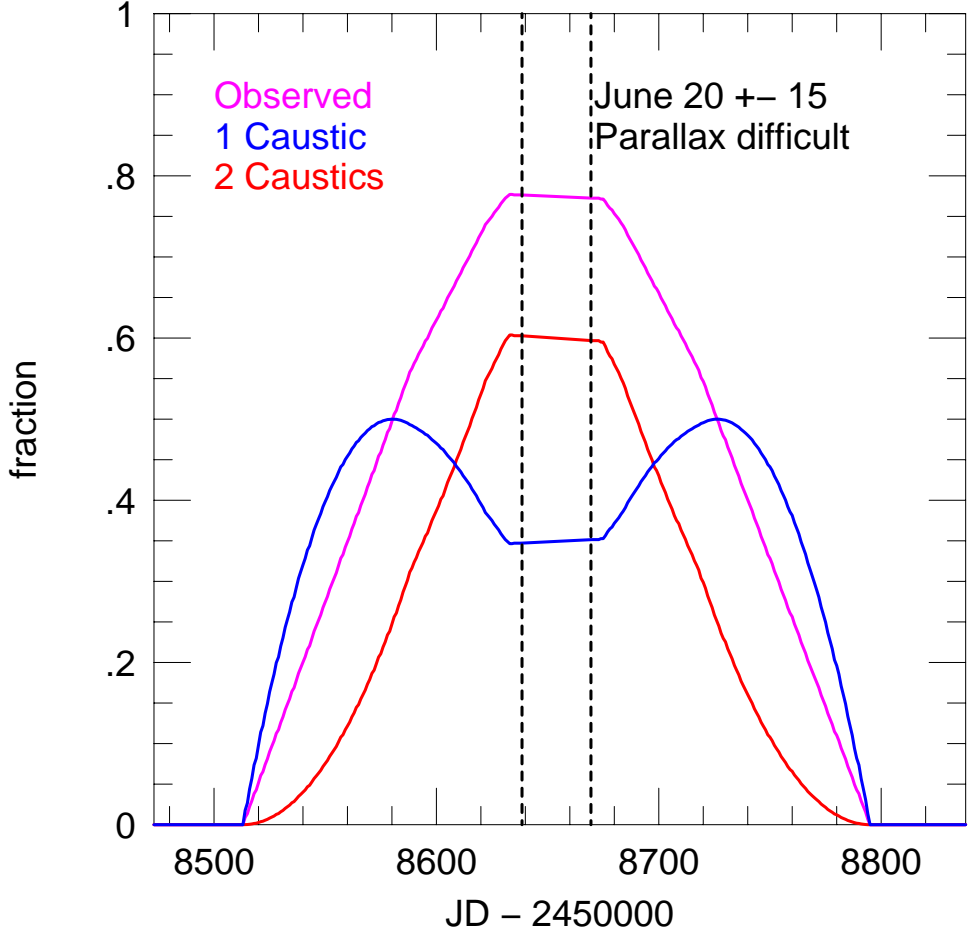}
\caption{Fraction of time that 
A: a given 1L1S event can be observed by some KMT observatory (magenta);
B: exactly one caustic of a given 2L1S event can be observed by some 
KMT observatory (blue); and
C: both caustics of a given 2L1S event can be observed by some KMT observatory
(red), assuming that KMT(A,S,C) are incapacitated by weather/Moon
(35,25,15)\% of the time.  Hence, in the parts of the season that
the red quadrangles in Figure~\ref{fig:kmt} do not overlap, the
magenta curve is simply 75\% of the time that a given day lies inside one
of these quadrangles.  When they do overlap, account is taken
of the fact that observations could take place at either observatory.
Then, $f_{\rm red}=f_{\rm magenta}^2$ and
$f_{\rm blue}=2(f_{\rm magenta}-f_{\rm red})$.  The black dashed lines indicate
the times of $D_\perp < 0.0025\,\au$ (when 2L1S $\pi_\e$ measurements are
very difficult) under the assumption that the satellite is exactly at L2.
In fact, L2 orbits have a minimum $24^\circ$ in-plane motion, so this
``dead zone'' will actually lie somewhere $\pm 25\,$days from where it
is shown.  However, this is a minor effect.  For 1L1S,
the range of ``difficult'' parallax measurements (not shown) is about
two times longer.
}
\label{fig:obs}
\end{figure}

\begin{figure}
\plotone{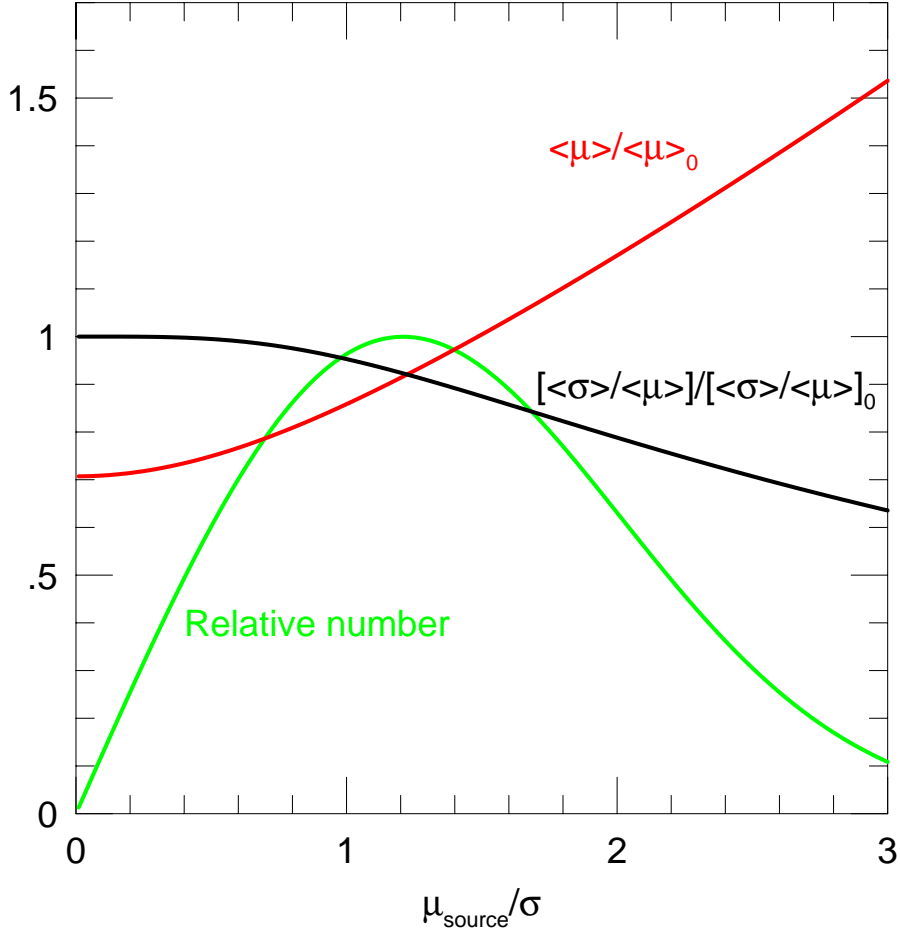}
\caption{\ch Effect of making a measurement of the source proper motion
in units of the 1-D source proper-motion dispersion.  The red
curve shows the expected proper motion relative to the case
of no measurement of $\bmu_s$, while the black curve shows a
similar comparison for the fractional error in the $\mu_\rel$ estimate.
Given Equation~(\ref{eqn:massest2}), these comparisons are exactly
the same for the mass estimate.  Hence, for example, if $\mu_s=0$, then
the mass estimate should be reduced by $\sqrt{2}$ but the fractional
error is exactly the same.  On the other hand, if $\mu_s=2\sigma$,
then the mean mass estimate is 17\% higher than the no-$\mu_s$-measurement
case, while the fractional error drops to 79\% of the no-$\mu_s$-measurement 
case.  The green curve shows the relative number of events with $\mu_s$
at various values.  This plot is constructed assuming that the
directional information from the parallax measurement is ignored.
}
\label{fig:muave}
\end{figure}

\end{document}